\documentclass{article}

\usepackage{amsmath}
\usepackage{amsfonts}
\usepackage{amssymb}

\usepackage{graphicx}

\begin{document}

\title{Bounding Lossy Compression using Lossless Codes at Reduced Precision}

\author{John Scoville}
\maketitle

\begin{abstract}
An alternative approach to two-part 'critical compression' is presented.  Whereas previous results were based on summing a lossless code at reduced precision with a lossy-compressed error or noise term, the present approach uses a similar lossless code at reduced precision to establish absolute bounds which constrain an arbitrary lossy data compression algorithm applied to the original data.
\end{abstract}

\section{Introduction}

Another possible implementation of critical data compression\cite{critical} compresses the critical bits of a data object losslessly, as before, while simultaneously compressing the entire object using lossy methods, as opposed to lossy coding only an error or residual value.  In principle, this results in the coding of redundant information.  In practice, however, the lossy coding step is often more effective when, for example, an entire image is compressed rather than just the truncated bits.  Encoding an entire data object tends to improve prediction in the lossy coder, while encoding truncated objects often leads to the high spatial frequencies which tend to be lost during lossy coding.  Such a redundant lossy coding of the original data often results in the most compact representation available, making this approach desirable for many applications relating to lossy coding.  This may not always be the case, for instance, when the desired representation is nearly lossless such a scheme may converge more slowly than one encoding truncated data.  On the other hand, however, this scheme is simple generally applicable to any type of lossy coding, whereas greater care must be taken when lossy coding truncated data (in video data, for instance) to avoid introducing high-frequency artifacts and noise to the data.  Furthermore, compressing and decompressing data using this approach potentially requires fewer operations than critically compressing data via truncation, as no normalization needs to be performed.

\section{The Algorithm}

Once the desired critical bit depth has been selected, the original data's precision is reduced from the original bit depth of $d$ to the critical bit depth of $n$ and the resulting object is coded losslessly, as before, while the original data is simultaneously coded using lossy compression.  As before, this operation may take place on a channel-by-channel basis with arbitrary parameters and with data having undergone an arbitrary set of transformations, e.g. color space rotations.

Upon decompression, the lossless code is used to establish exact upper and lower bounds for lossy compression.  The lossless code establishes a lower bound since the truncated values never exceed the original values.  Likewise, since $2^{d-n}-1$ is the largest quantity which could have be truncated, adding this to the reduced-precision data produces an upper bound on any value which could be truncated to produce the losslessly coded reduced-precision data.

The decompression scheme is as follows: If the $n$ leading bits of the value predicted by the lossy code match the lossless code, the value predicted by the lossy code is returned as the decompressed datum.  If the $n$ leading bits predicted by the lossy code are less than the value coded by the lossless code, the value of the lossless code is returned as the decompressed datum since it is a lower bound on the original value.  Otherwise, if the $n$ leading bits predicted by the lossy code are greater than the the value coded by the lossless code, the value coded by the lossless code is increased by $2^{d-n}-1$ and returned as the decompressed datum since this is an upper bound on the value of the original data.

\section{Example}

As an example of this approach, we will consider the critical compression of another test image, the 8-bit (per channel) RGB color 'fireworks' image from 'The New Test Images', available at imagecompression.info.  The image is 3136 by 2152 pixels, which is 20,246,016 bytes of raw data.  We will compress the image in YCC space, with the luminance channel being critically compressed at a bit depth of 4 and chrominance data is taken from the lossy compression.  Two objects are stored, one is a JPEG2000 compressed version of the original RGB color space image (which uses a YCC color space internally) whose compression ratio is 1000:1 and the other is a PAQ-compressed lossless representation of a luminance (Y) channel whose precision has been reduced from an 8-bit (256 shade) grayscale image to a 4-bit (16 shade) grayscale image.  The JPEG2000 representation of the original image may be seen in figure 1, and the lossless representation of a 4-bit luminance channel derived from the original image in figure 2.

\begin{center}
  \includegraphics[width=4.7in]{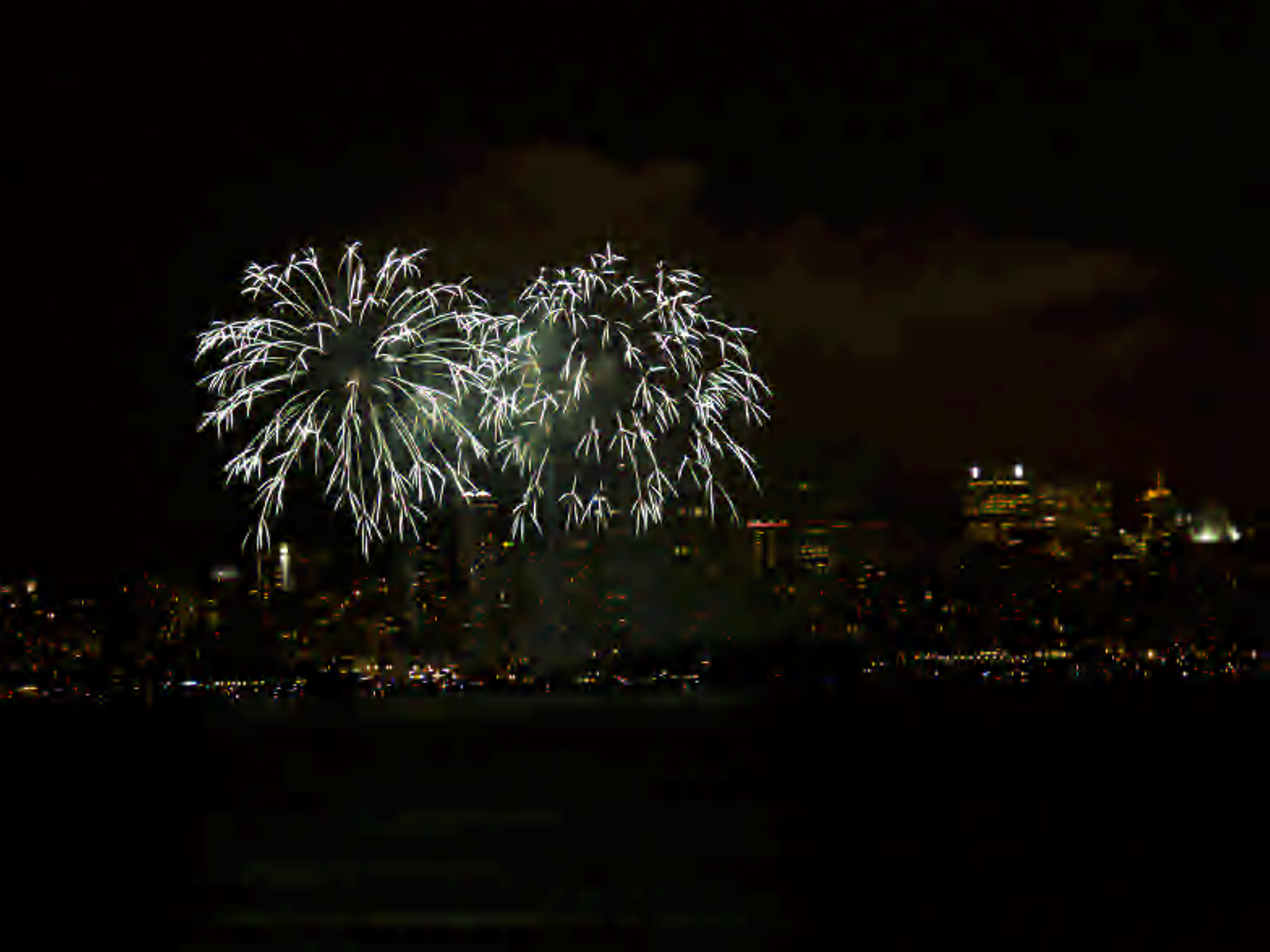}\\
Figure 1\\
\end{center}

\begin{center}
  \includegraphics[width=4.7in]{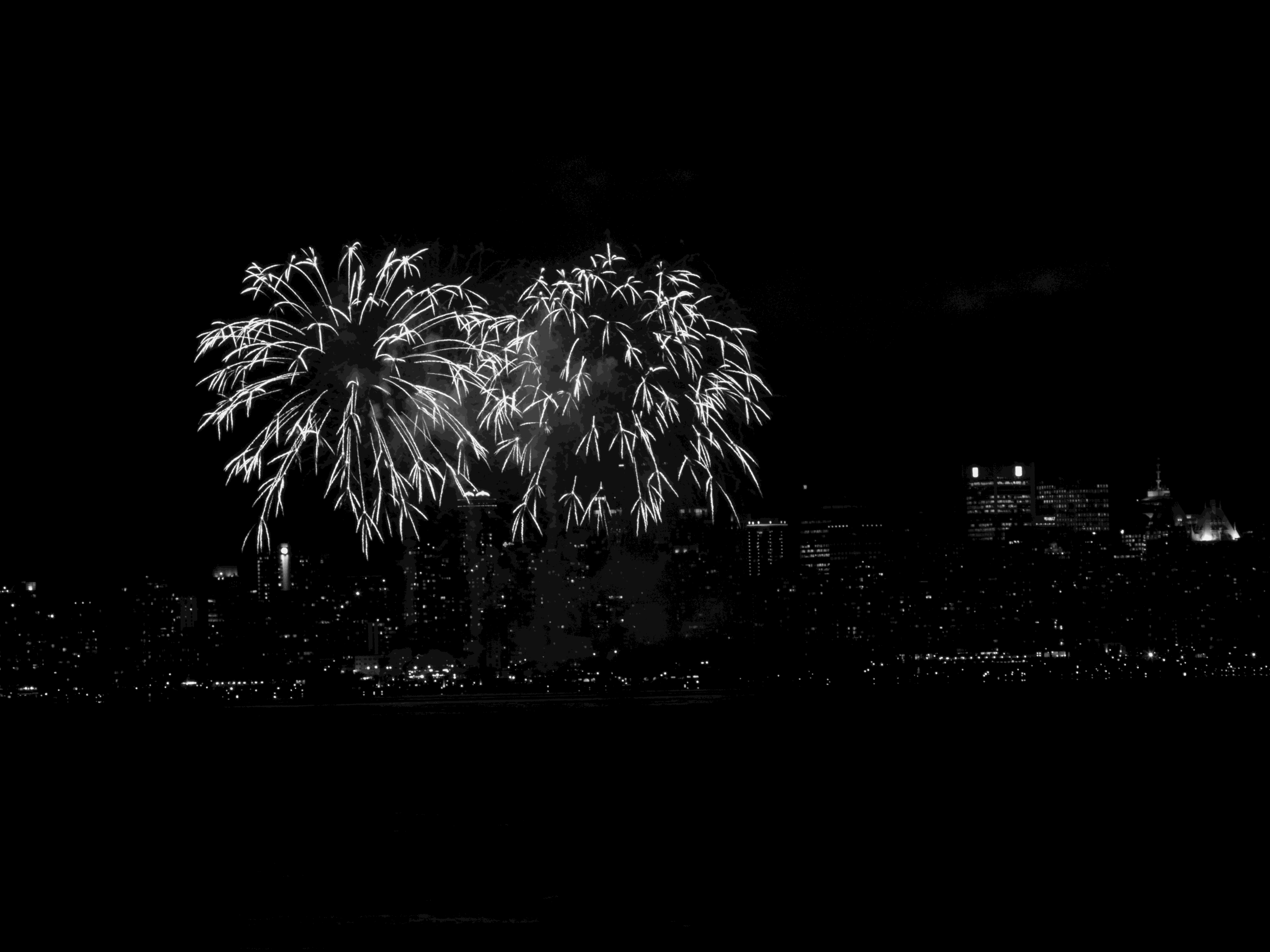}\\
Figure 2\\
\end{center}

During decompression, the luminance value implied by the color of each pixel in the JPEG2000 channel is compared to the 4-bit luminance channel. The largest possible truncation error in the 4-bit luminance channel is 15, so the luminance should be between the value coded by the 4-bit lossless channel and this value increased by fifteen.  If the luminance predicted by the lossy JPEG2000 code falls within this range, then the color predicted by JPEG2000 is used for the pixel.  If the JPEG2000-predicted luminance falls below this range, then the value coded by the 4-bit lossless channel is used for the luminance, being combined with the chrominance values implied by the JPEG2000-predicted color value before being rotated back into the RGB color space.  If the JPEG2000-predicted luminance falls above the allowed range, then the value coded by the 4-bit lossless channel is increased by fifteen before used for the luminance and combined with the chrominance values implied by the JPEG2000-predicted color and being rotated back into the RGB color space.  The decompressed image resulting from this coding and decoding scheme may be seen below.  Examination of this image reveals that the absolute bounds obtained from lossless brightness values have greatly increased the contrast of the resulting picture, as compared to the JPEG2000 representation.

\begin{center}
  \includegraphics[width=4.7in]{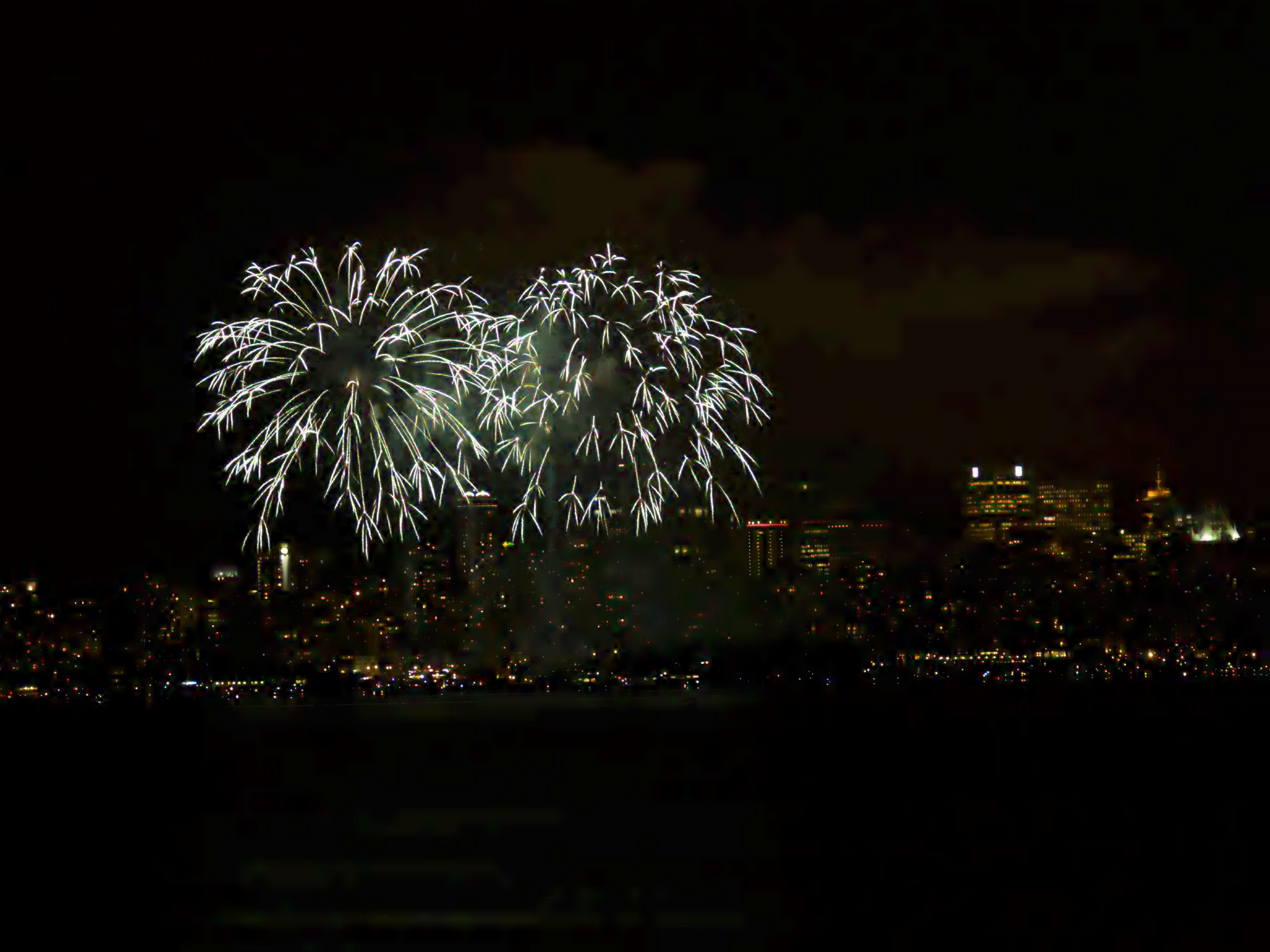}\\
Figure 3\\
\end{center}

\section{Disclosures}
This research was funded entirely by the author, John Scoville, and the method described is part of a pending patent application.

\providecommand{\bysame}{\leavevmode\hbox to3em{\hrulefill}\thinspace}
\providecommand{\MR}{\relax\ifhmode\unskip\space\fi MR }
\providecommand{\MRhref}[2]{%
  \href{http://www.ams.org/mathscinet-getitem?mr=#1}{#2}
}
\providecommand{\href}[2]{#2}

\end{document}